# Background Independent Quantum Mechanics, Metric of Quantum States, and Gravity: A Comprehensive Perspective

Aalok

Department of Physics, University of Rajasthan, Jaipur 302004, India,
and Jaipur Engineering College and Research Centre (JECRC), Jaipur 303905, India.

**Abstract**

This paper presents a comprehensive perspective of the metric of quantum states with a focus on geometry in the background independent quantum mechanics. We also explore the possibilities of geometrical formulations of quantum mechanics beyond the quantum state space and Kähler manifold. The metric of quantum states in the classical configuration space with the pseudo-Riemannian signature and its possible applications are explored. On contrary to the common perception that a metric for quantum state can yield a natural metric in the configuration space when the limit $\hbar \to 0$, we obtain the metric of quantum states in the configuration space without imposing the limiting condition $\hbar \to 0$. Here Planck's constant $\hbar$ is absorbed in the quantity like Bohr radii $\frac{1}{2mZ\alpha} \sim a_0$. While exploring the metric structures associated with Hydrogen like atom, we witness another interesting finding that the invariant lengths appear in the multiple of Bohr's radii as: $ds^2 = a_0^2 (\nabla \Psi)^2$.

___________________________

*E-mail*: aalok@uniraj.ernet.in

# 1. INTRODUCTION

In the light of recent studies of geometry [1-5] of the quantum state space, the need and call for further extension of standard geometric quantum mechanics is irresistible. And thus an intensive follow up will be academically rewarding. Researchers studying gravity have also shown considerable interest in the geometric structures in quantum mechanics in general and projective Hilbert space in specific [2-7].

Classical mechanics has deep roots in (symplectic) geometry while quantum mechanics is essentially algebraic. However, one can recast quantum mechanics in a geometric language, which brings out the similarities and differences between two theories [6]. The idea is to pass from the Hilbert space to the space of rays, which is the "true" space of states of quantum mechanics. The space of rays- or the projective Hilbert space is in particular, a symplectic manifold, which happens to be equipped with a Kähler structure. Regarding it as a symplectic manifold, one can repeat the familiar constructions of classical mechanics. Precisely, one of our motifs in this paper is to be able to repeat the familiar constructions of classical mechanics in quantum geometric formalism. The present paper begins with the generalized formalism in quantum geometry discussed recently [1], and attempts to project a broad perspective based on it.

The *distance* on the projective Hilbert space is defined in terms of metric, called the metric of the ray space or the projective Hilbert space $\mathscr{P}$, is given by the following expression in Dirac's notation:

$$ds^2 = 4\left(1 - \left|\langle \Psi_1 | \Psi_2 \rangle\right|^2\right) \equiv 4\left(\langle d\Psi | d\Psi \rangle - \langle d\Psi | \Psi \rangle \langle \Psi | d\Psi \rangle\right). \tag{1}$$

This can be regarded as an alternative definition of the Fubini-Study metric, valid for an infinite dimensional $\mathscr{H}$.

The metric in the ray space is now being referred by physicists as the background independent and space-time independent structure, which can play an important role in the construction of a potential "theory of quantum gravity". The demand of background independence in a quantum theory of gravity calls for an extension of standard geometric quantum mechanics [2-4]. The metric structure in the projective Hilbert space is treated as background independent and space-time independent geometric structure. It is important insight which can be the springboard for our proposed background independent generalization of standard quantum mechanics. For a generalized coherent state, the FS metric reduces to the metric on the corresponding group manifold [2-3]. Thus, in the wake of ongoing work in the field of quantum geometric formulation, the work in the present paper may prove to be very useful. The probabilistic (statistical) interpretation of QM is thus hidden in the metric properties of $\mathscr{P}(\mathscr{H})$. The unitary time evolution is related to the metrical structure [2-3] with Schrödinger's equation in the guise of a geodesic equation on $CP(N)$. The metric in equation (1) is real and positive definite [8-10]. We cannot expect a metric with the signature of Minkowski space in the study of the metric of quantum state space, as the metric of quantum state space is in the projective Hilbert space and therefore it is always positive definite. However, we can define the metric of quantum states in the configuration space, but such a metric need not always be positive definite. To be precise, the metric of quantum state space is a metric on the underlying manifold which the quantum states form or belong to, and therefore, it is different from the metric of space-time or any other metric associated with the quantum states.

A quantum state in the Hilbert space corresponds to a point in the projective Hilbert space, by means of projections. Inverse of these projections are known as fibers. And two points in the projective Hilbert space can lie on a line which stands for neighborhood in topological sense provided the corresponding two states in the Hilbert space are connected by means of invariance under local gauge transformations. The basic objective behind formulation of the metric of quantum state space was to seek invariance in the quantum evolution under the local gauge transformations [8-13]. One can verify this fact from the equation (1); where, there are two parts in the expression of metric coefficient $g_{\mu\nu}$, such that whenever the first part picks up an additional term due to local gauge transformation, it gets cancelled by a similar extra term picked up by the second part. Thus, the metric of quantum state space is invariant under the local gauge transformations in addition to the invariance under coordinate transformations. As rightly pointed out by Minic and Tze, everything we know about quantum mechanics [2-4] is in fact contained in the geometry of $CP(N)$. Entanglements come from the embeddings of the products of two complex projective spaces in a higher dimensional one; geometric phase stem from the symplectic structure of $CP(N)$, quantum logic, algebraic approaches to quantum mechanics etc, are all contained in the geometric and symplectic structure of complex projective spaces [2-4]. While we only consider here the finite dimensional case, the same geometric approach is extendible to generic infinite dimensional quantum mechanical systems, including field theory. Finally, the following three lemmas summarize this discussion as:

(i) The Fubini-Study metric as given in the equation (1) and (3) in the limit $\hbar \to 0$ becomes a spatial metric, provided the configuration space for the quantum system under consideration is space-time. For example, if we consider a particle moving in 3-dimensional Euclidean space, then the quantum metric for the Gaussian coherent state

$$\Psi_l(x) \sim \exp\left(-\frac{(\vec{x}-\vec{l})^2}{\delta l^2}\right)$$ yields the natural metric in the configuration space, in the limit $\hbar \to 0$, becomes $ds^2 = \frac{d\vec{l}^2}{\delta l^2}$. \hfill (2)

(ii) Similarly, the time parameter of the evolution equation can be related to the quantum metric *via*

$$\hbar ds = \Delta E dt, \quad \Delta E^2 \equiv \langle \Psi | H^2 | \Psi \rangle - \langle \Psi | H | \Psi \rangle^2. \tag{3}$$

(iii) Finally, the Schrödinger equation can be viewed as a geodesic equation on a $CP(N) = \frac{U(N+1)}{U(N) \times U(1)}$ as:

$$\frac{du^a}{ds} + \Gamma^a_{bc} u^b u^c = \frac{1}{2\Delta E} Tr(HF^a_b) u^b. \tag{4}$$

Here $u^a = \frac{dz^a}{ds}$ where $z^a$ denote the complex coordinates on $CP(N)$, $\Gamma^a_{bc}$ is the connection obtained from the Fubini-Study metric, and $F_{ab}$ is the canonical curvature 2-form valued in the holonomy gauge group $U(N) \times U(1)$. Here, Hilbert space is $N+1$ dimensional and the projective Hilbert space has dimenssions $N$. Furthermore, there is enlarged vision of these symmetries explored recently which is discussed in the section 2.2 of this paper in the context of Background independent quantum mechanics (BIQM).

However, on contrary to the common perception that a metric for quantum state can yield a natural metric in the configuration space when the limit $\hbar \to 0$, we find the metric of quantum states in the configuration space without imposing the limiting condition $\hbar \rightarrow 0$. The Planck's constant $\hbar$ is absorbed in the quantity like Bohr radii $\frac{1}{2mZ\alpha} \sim a_0$, where $\alpha = \frac{e^2}{\hbar c}$ is fine structure constant and $Z$ is corresponding atomic number.

The motivation behind our formulation in this paper is two fold: firstly, to explore a wider perspective for the generalised definition of the metric of quantum states, and secondly to think beyond the quantum state space in search of pseudo-Riemannian structures by exploring the metric of quantum states in the configuration space with the signature of Lorentzian or a Minkowskian metric and its possible applications. Also, we discuss the metric of quantum states in the configuration space and its invariance under coordinate transformations and the Lorentz' transformations.

## 2. THE METRIC OF QUANTUM STATES: GENERALISED DEFINITION AND SPACE-TIME INDEPENDENT METRICS

The generalised definition of the metric of quantum states was laid down recently by Aalok *et al* [1], using first principles of differential geometry. The invariant corresponding to this generalised formulation of metric was prescribed as:

$$ds^2 = |\nabla\Psi|^2 = (\nabla_\mu \Psi)(\nabla_\nu \Psi) dx^\mu dx^\nu. \qquad (5)$$

The metric tensor $g_{\mu\nu}$ for this invariant can be given as:

$$g_{\mu\nu} = \text{Re}[(\nabla_\mu \Psi)(\nabla_\nu \Psi)]. \qquad (6)$$

Alternatively, one can also write the symmetric tensor $g_{\mu\nu}$ as

$$g_{\mu\nu} = (\nabla_\mu \Psi)^*(\nabla_\nu \Psi) = \frac{1}{2}\left[(\nabla_\mu \Psi)^*(\nabla_\nu \Psi) + ((\nabla_\mu \Psi)^*(\nabla_\nu \Psi))^*\right]$$

$$= \frac{1}{2}\left[(\nabla_\mu \Psi)^*(\nabla_\nu \Psi) + (\nabla_\nu \Psi)^*(\nabla_\mu \Psi)\right]. \qquad (7)$$

We find that this generalized definition satisfies all geometrical requirements of metric structure [1].

Following this generalized definition, the metric of quantum state space, and the metric of quantum states in the configuration space is deduced. We also illustrate some examples on it.

## 2.1. THE METRIC OF QUANTUM STATE SPACE

From the generalized definition discussed here, we reproduce the expression of the metric of quantum state space. We consider a quantum state $\Psi \equiv \Psi\{\lambda\}$, $\forall \Psi \in \mathcal{H}$, and the corresponding covariant derivative for the quantum states [8] is given by:

$$\nabla^\lambda \Psi \equiv \left|\frac{d\Psi}{d\lambda}\right\rangle + \left\langle \Psi \left| \frac{d\Psi}{d\lambda} \right\rangle |\Psi\rangle. \qquad (8)$$

Here, $\lambda$ in equation (8) could be local co-ordinates on $\mathcal{P}$. Applying this covariant derivative to the definition of metric in eq. (7) we obtain the desired metric coefficients:

$$g_{\lambda\lambda} = \left[(\nabla_\lambda \Psi)^*(\nabla_\lambda \Psi)\right] = \frac{1}{2}\left[(\nabla_\lambda \Psi)^*(\nabla_\lambda \Psi) + \left((\nabla_\lambda \Psi)^*(\nabla_\lambda \Psi)\right)^*\right]$$

$$= \left[\left(\left\langle \frac{\partial \Psi}{\partial \lambda}\right| - \langle\Psi|\left\langle\frac{\partial \Psi}{\partial \lambda}\middle|\Psi\right\rangle\right)\left(\left|\frac{\partial \Psi}{\partial \lambda}\right\rangle - \left\langle\Psi\middle|\frac{\partial \Psi}{\partial \lambda}\right\rangle|\Psi\rangle\right)\right],$$

$$= \left\langle \frac{\partial \Psi}{\partial \lambda}\middle|\frac{\partial \Psi}{\partial \lambda}\right\rangle - \left\langle \frac{\partial \Psi}{\partial \lambda}\middle|\Psi\right\rangle\left\langle\Psi\middle|\frac{\partial \Psi}{\partial \lambda}\right\rangle - \left\langle \frac{\partial \Psi}{\partial \lambda}\middle|\Psi\right\rangle\left\langle\Psi\middle|\frac{\partial \Psi}{\partial \lambda}\right\rangle + \left\langle \frac{\partial \Psi}{\partial \lambda}\middle|\Psi\right\rangle\left\langle\Psi\middle|\frac{\partial \Psi}{\partial \lambda}\right\rangle\langle\Psi|\Psi\rangle.$$

Which gives $g_{\lambda\lambda} = \left[\left\langle \frac{\partial \Psi}{\partial \lambda}\middle|\frac{\partial \Psi}{\partial \lambda}\right\rangle - \left\langle \frac{\partial \Psi}{\partial \lambda}\middle|\Psi\right\rangle\left\langle\Psi\middle|\frac{\partial \Psi}{\partial \lambda}\right\rangle\right].$ (9)

Also, we can write it in a generalized way as:

$$g_{\mu\nu} = \left[\left\langle \frac{\partial \Psi}{\partial x_\mu}\middle|\frac{\partial \Psi}{\partial x_\nu}\right\rangle - \left\langle \frac{\partial \Psi}{\partial x_\mu}\middle|\Psi\right\rangle\left\langle\Psi\middle|\frac{\partial \Psi}{\partial x_\nu}\right\rangle\right]. \qquad (10)$$

This is same as the metric of quantum state space, was also formulated [1, 8-13] for the real local coordinates $x^\mu$. But this metric is no more on Kähler manifold. If the metric of quantum states is defined with local co-ordinates that are not complex, it lies on the base manifold with Riemannian character, and the local gauge group $GL(n,R)$ is also admissible, where $n$ is the dimensionality of the space-time.

If we consider the relativistic evolution of quantum states by Klein-Gordon equation as follow:

$$-\nabla^\mu \nabla_\mu \Psi = \frac{m_0^2 c^2}{\hbar^2} \Psi . \tag{11}$$

We immediately realise a covariant and invariant quantity resulting from it [1]:

$$-\Psi^* \nabla^\mu \nabla_\mu \Psi = \frac{m_0^2 c^2}{\hbar^2} \Psi^* \Psi . \tag{12}$$

This expression is covariant and also invariant under local gauge transformations. Being inspired by the covariance and the invariance of this expression, one can formulate a metric [1] for quantum states from it as follow:

$$\Psi^* \nabla_\mu \nabla_\nu \Psi = \frac{1}{2}\left[(\Psi^* \nabla_\mu \nabla_\nu \Psi) + (\Psi^* \nabla_\mu \nabla_\nu \Psi)^*\right]; \tag{13}$$

$$= -[\langle \partial_\mu \Psi | \partial_\nu \Psi \rangle - \langle \partial_\mu \Psi | \Psi \rangle \langle \Psi | \partial_\nu \Psi \rangle]. \tag{14}$$

Thus, we find that the above invariant is the well-known expression of the metric of quantum states as:

$$g_{\mu\nu} = [\langle \partial_\mu \Psi | \partial_\nu \Psi \rangle - \langle \partial_\mu \Psi | \Psi \rangle \langle \Psi | \partial_\nu \Psi \rangle] . \tag{15}$$

The metric of quantum state space has been identified as background independent (BI) metric structure [2-5]. However, by appearance itself the invariance of the geometric structure in equation (15) is apparent, irrespective of the choice of state function.

In the context of complex projective space $CP$, due to $Diff(\infty, C)$ symmetry, the "coordinates" $Z^a$ while representing quantum states, make no sense physically, only quantum events do, which is the quantum counterpart of the corresponding statement on the meaning of space-time events in General Relativity (GR). Probability is generalised and given by the notation of diffeomorphism invariant distance in the space of quantum configurations. The dynamical equation is a geodesic equation in this space. Time, the evolution parameter in the generalised Schrödinger equation, is yet not global and is given in terms of the invariant distance. The basic point as threshold of the background independent quantum mechanics (BIQM) is to notice that the evolution equation (the generalised Schrödinger equation) as a geodesic equation, can be derived from an Einstein-like equation with the energy-momentum tensor determined by the holonomic non-abelian field strength $F_{ab}$ of the $Diff(\infty - 1, C) \times Diff(1, C)$ type and the interpretation of the Hamiltonian as a charge.

Such an extrapolation is logical since $CP(N)$ is an Einstein space, and its metric obeys Einstein's equation with a positive cosmological constant given by:

$$R_{ab} - \frac{1}{2} R g_{ab} - \Lambda g_{ab} = 0. \tag{16}$$

The diffeomorphism invariance of the new phase space suggests the following dynamical scheme for the (BIQM) as: $R_{ab} - \frac{1}{2} R g_{ab} - \Lambda g_{ab} = T_{ab},$ (17) with $T_{ab}$ be given as above.

Furthermore, $\nabla_a F^{ab} = \frac{1}{2\Delta E} H u^b$. (18)

The last two equations imply *via* the Bianchi identity, a conserved energy-momentum tensor: $\nabla_a T^{ab} = 0$. (19)

This taken together with the conserved "current" as: $j^b = \frac{1}{2\Delta E} H u^b$, and $\nabla_a j^a = 0$; (20)

implies the generalised geodesic Schrödinger equation. Thus equation (17) and (18), being a closed system of equations for the metric and symplectic structure do not depend on the Hamiltonian, which is the case in ordinary quantum mechanics. By imposing the conditions of homogeneity and isotropy on the metric by means of number of Killing vectors, the usual quantum mechanics can be recovered [2-5, 6]. And this limit does not affect the geodesic equation

$$\frac{du^a}{ds} + \Gamma^a_{bc} u^b u^c = \frac{1}{2\Delta E} Tr(HF^a_b) u^b,$$ (21)

due to the relation $\hbar d\tau = 2\Delta E dt$. (22)

The reformulation of the geometric QM in this background independent setting gives us lot of new insights. The utility of the BIQM formalism is that gravity embeds into quantum mechanics with the requirement that the kinematical structure must remain compatible with the generalized dynamical structure under deformation. The requirement of diffeomorphism invariance places stringent constraints on the quantum geometry. We must have a strictly (i.e. non-integrable) almost complex structure on the generalized space of quantum events. The symmetries as described by the quotient set $CP(N) = \frac{U(N+1)}{U(N) \times U(1)}$, have limitations. In an extended framework of geometric quantum mechanics the invariance of the metric structure had been suggested [2-5] for $CP(\infty)$ as $\dfrac{Diff(\infty, C)}{Diff(\infty-1, C) \times Diff(1, C)}$. By insisting on the diffeomorphism invariance in the state space and on preserving the desirable complex projective properties of Cartan's rank 1 symmetric spaces such as $CP(N)$, we arrive at the ensuing coset state space $\dfrac{Diff(\infty, C)}{Diff(\infty-1, C) \times Diff(1, C)}$ as the minimal phase space candidate for a background independent quantum mechanics (BIQM). But, this does not seem to guarantee an almost complex structure [5]. Thus the only alternative seem to satisfy the almost complex structure is the Grassmannian. By the correspondence principle, the generalized quantum geometry must locally recover the canonical quantum theory encapsulated in $\mathscr{P}$ ($\mathcal{N}$) and also allows for mutually compatible metric and symplectic structure, supplies the framework for the dynamical extension of the canonical quantum theory.

The Grassmannian: $Gr(C^{n+1}) = Diff(C^{n+1}) / Diff\left(C^{n+1}, C^n \times \{0\}\right)$. (23)

In the limit $n \to \infty$ limit satisfies the necessary conditions [5]. This space is generalization of $\mathscr{P}$ ($\mathcal{N}$). The Grassmannian is a gauged version of complex projective space, which is the geometric realization of quantum mechanics. The utility of this formalism is that gravity embeds into quantum mechanics with the requirement that the kinematical structure must remain compatible with the generalization dynamical structure under deformation. The quantum symplectic and metric structure, and therefore the almost complex structure, are themselves fully dynamical.

## 2.2. THE METRIC OF QUANTUM STATES IN A CONFIGURATION SPACE

In this exercise we explore the possibilities beyond the geometry of projective Hilbert space and Kähler manifold. Consequently, we aim to get metric of quantum states with the classical nature.

In the formalism of geometric quantum mechanics, coordinates are not meaningful. On a Kähler manifold in the quantum state space, invariance under the local gauge transformations is same as invariance under the coordinate transformations. This is due to the reason that in the quantum state space, quantum states themselves could play the role of coordinates.

On the other hand, a metric with classical nature does not admit invariance under the local gauge transformations. And for which, the invariance under the coordinate transformations is enough. However, if we compromise by not retaining the invariance under local gauge transformations, and still ensuring the invariance under coordinate transformations, we can obtain a metric with classical nature from a generalised definition of metric.

Thus, we explore the possibility of a scenario where invariance under the local gauge transformations may be lost but invariance under coordinates is still retained.

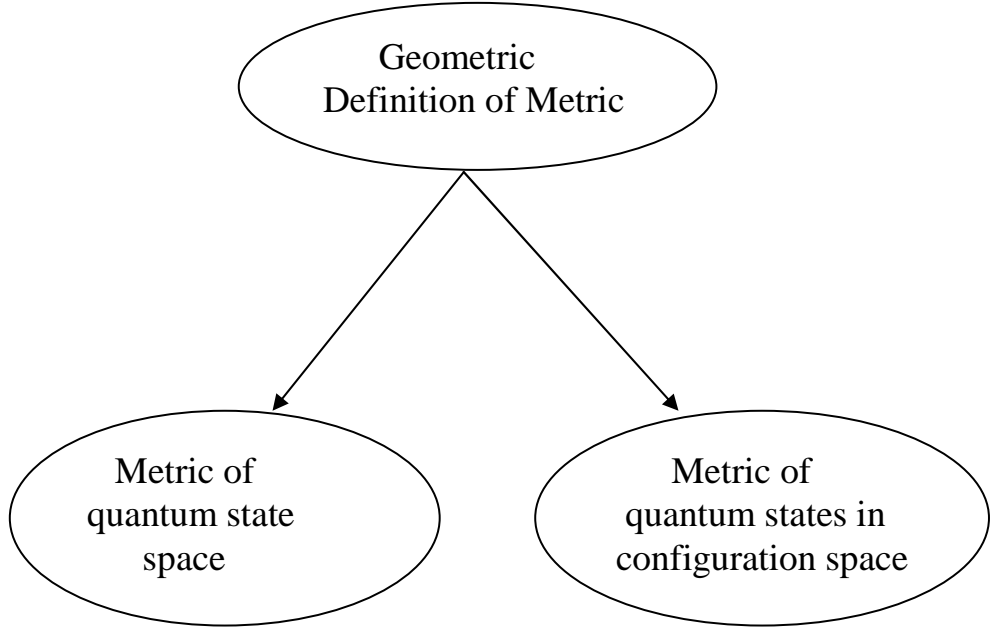

**Fig. 1.** A scheme of metric formalism on different manifolds

The definition of the metric tensor in (6) and (7) involves only first order derivatives, thus even if we use ordinary partial derivatives instead of the covariant derivative defined in eq. (8), the metric properties of $g_{\mu\nu}$ remain unaffected. Also, even if we do not apply the complex conjugation, and consider only the real part of eq. (6), we still retain the metric structure. However, for such a metric positive-definiteness is no more assured, as it is not the metric of quantum state space and no pull back metric exists for this metric. Moreover, this is metric in the configuration space, and the nature and signature of the metric will depend upon the choice of wave function. We redefine our metric as:

$$ds^2 = \text{Re } (\nabla\Psi)^2 = \text{Real Part}\left[(\nabla_\mu\Psi)(\nabla_\nu\Psi)\right]dx^\mu dx^\nu, \tag{24}$$

such that, $g_{\mu\nu} = \text{Real Part}\left[(\nabla_\mu\Psi)(\nabla_\nu\Psi)\right] = \text{Real Part}\left[\left(\dfrac{\partial\Psi}{\partial x_\mu}\right)\left(\dfrac{\partial\Psi}{\partial x_\nu}\right)\right]. \tag{25}$

It has been shown [1] that the *ds* being differential form guarantees invariance of this metric under the coordinate transformations, and the quantity $g_{\mu\nu} = \left(\frac{\partial \Psi}{\partial x_\mu}\right)\left(\frac{\partial \Psi}{\partial x_\nu}\right)$ is a transformable quantity.

Though, quantum states live in Hilbert space, they represent physical states and do depend on the parameters of the physical configuration space. Thus, it is just not possible that they do not affect the configuration space in which they describe physical systems. Thus one can say that metric of quantum states in the configuration space being discussed here is the imprint of the quantum states which they leave on the configuration space. This is precisely the essence of metric in the configuration space. We cannot say anything further about the physical significance of this metric, unless we choose a specific physical function.

To avoid confusions, we clarify that the metric on configuration space is not at all being deduced from the Fubini-Study metric. We have a generalized geometric structure in the beginning, from which we deduce the metric of quantum state space as well as metric on the configuration space. One may be surprised, "How do we get two different metric structures from a generalized definition?" Answer is simple! The coordinates used in case of metric of quantum states in the quantum state space, are real local coordinates on the manifold of the quantum states in the projective Hilbert space $\mathscr{P}$. Where as, in case of metric in configuration space, the coordinates used are the coordinates on space-time. Also, one could notice the reasons for invariance of the metric of quantum states in ray space under local gauge transformations. The 'connection'- $\left\langle \Psi \middle| \frac{\partial \Psi}{\partial \lambda} \right\rangle$ sitting inside the covariant derivative $\nabla^\lambda \Psi \equiv \left|\frac{d\Psi}{d\lambda}\right\rangle + \left\langle\Psi\left|\frac{d\Psi}{d\lambda}\right\rangle\right|\Psi\rangle$, and having rooted its feet in local coordinates, always keeps connecting the initial state with the final state. This results into the invariance of the metric of the ray space under local gauge transformations, which is precisely the essence of the metric formulation in ray space. In case of metric in the configuration space it does not happen, and metric remains invariant only under coordinate transformations. It should be noticed that if the metric of quantum states is defined in the configuration space with the space-time co-ordinates, the base manifold **M** on which it lies, carries a (pseudo) Riemannian metric as well, and the tetrad can naturally be chosen to bring the metric $g_{\mu\nu}$ to a diagonal Minkowski form, and then the Lorentz' group $SO(3,1)$ appears as a local gauge group.

We now illustrate some examples, to show how different metric structures with different signatures could be obtained.

## 3. METRIC CORRESPONDING TO HYDROGEN LIKE ATOM

To illustrate an example, we describe the metric structure corresponding to the Hydrogen atom wave function. For this, we consider the eigen functions $\Psi_{100}, \Psi_{200}, \Psi_{210}, \Psi_{211}$ of Hydrogen atom; generally represented by $\Psi_{nlm}$, where *n*, *l*, and *m* are principal, azimuthal and magnetic quantum numbers respectively.

Firstly, we choose the wave function $\Psi_{100}$ of the Hydrogen like atom:

$$\Psi_{100} = \left(\frac{1}{\sqrt{\pi}}\right)\left(\frac{1}{a_0}\right)^{3/2}\left(e^{-\frac{r}{a_0}}\right)\left(e^{-i\frac{W_1}{\hbar}t}\right). \tag{26}$$

Here $a_0 = \dfrac{\hbar^2}{\mu e^2} = .529 \times 10^{-8} cm$, is the Bohr's radius, and $W_1 = -\dfrac{\mu e^4}{2\hbar^2} = -2.15 \times 10^{-11} ergs$, is the lowest energy level of Hydrogen atom.

For brevity, we substitute $\dfrac{1}{\sqrt{\pi}}\left(\dfrac{1}{a_0}\right)^{3/2} = C_0$ and $\dfrac{W_1}{\hbar} = \omega_0$, such that the above wave function reduces to a simpler form: $\Psi_{100} = C_0\left(e^{-\dfrac{r}{a_0}}\right)\left(e^{-i\omega_0 t}\right)$. (27)

While working in the orthogonal co-ordinates the off-diagonal terms of the line element vanish, and the diagonal metric coefficients corresponding to the quantum state $\Psi \equiv \Psi(r,t)$, $\forall \Psi \in \mathscr{H}$ are given by:

$$g_{rr} = \text{Re}\left[\left(\dfrac{\partial \Psi}{\partial r}\right)\left(\dfrac{\partial \Psi}{\partial r}\right)\right] = \left[\left(\dfrac{C_0}{a_0}\right)^2\left(e^{-\dfrac{2r}{a_0}}\right)\cos 2\omega_0 t\right], \tag{28}$$

and $g_{tt} = \text{Re}\left[\left(\dfrac{\partial \Psi}{\partial t}\right)\left(\dfrac{\partial \Psi}{\partial t}\right)\right] = -\left[(C_0\omega_0)^2\left(e^{-\dfrac{2r}{a_0}}\right)\cos 2\omega_0 t\right]$. (29)

The corresponding invariant $ds$ appears as:

$$ds^2 = \left[\left(\dfrac{C_0}{a_0}\right)^2\left(e^{-\dfrac{2r}{a_0}}\right)\cos 2\omega_0 t\right]dr^2 - \left[(C_0\omega_0)^2\left(e^{-\dfrac{2r}{a_0}}\right)\cos 2\omega_0 t\right]dt^2. \tag{30}$$

We now describe metric for the wave function $\Psi_{200}$ of the Hydrogen atom:

$$\Psi_{200} = \left(\dfrac{1}{4\sqrt{2\pi}}\right)\left(\dfrac{1}{a_0}\right)^{3/2}\left(2 - \dfrac{r}{a_0}\right)\left(e^{-\dfrac{r}{2a_0}}\right)\left(e^{-i\dfrac{W_2}{\hbar}t}\right). \tag{31}$$

On substituting $\dfrac{1}{4\sqrt{2\pi}}\left(\dfrac{1}{a_0}\right)^{3/2}=C$ and $\dfrac{W_2}{\hbar}=\dfrac{W_1}{4\hbar}=\omega$, the above wave function reduces to a simpler form: $\Psi_{200}=C\left(e^{-\dfrac{r}{2a_0}}\right)\left(2-\dfrac{r}{a_0}\right)\left(e^{-i\omega t}\right)$.

And the metric structure corresponding to this quantum state $\Psi \equiv \Psi(r,t)$, is given by:

$$ds^2=\left[4\left(\dfrac{C}{a_0}\right)^2\left(e^{-\dfrac{r}{a_0}}\right)\cos 2\omega t\right]dr^2-\left[(C\omega)^2\left(2-\dfrac{r}{a_0}\right)^2\left(e^{-\dfrac{r}{a_0}}\right)\cos 2\omega t\right]dt^2. \tag{32}$$

Further we describe metric for the wave function $\Psi_{210}$ of the Hydrogen atom:

$$\Psi_{210}=\left(\dfrac{1}{4\sqrt{2\pi}}\right)\left(\dfrac{1}{a_0}\right)^{3/2}\left(\dfrac{r}{a_0}\right)(\cos\theta)\left(e^{-\dfrac{r}{2a_0}}\right)\left(e^{-i\dfrac{W_2}{\hbar}t}\right)=C\left(\dfrac{r}{a_0}\right)(\cos\theta)\left(e^{-\dfrac{r}{2a_0}}\right)\left(e^{-i\omega t}\right). \tag{33}$$

However, we can consider an un-normalized dimensionless form of this wave function for simplicity as: $\tilde{\Psi}=\left(\dfrac{r}{a_0}\right)(\cos\theta)\left(e^{-\dfrac{r}{2a_0}}\right)\left(e^{-i\omega t}\right)$. And redefine another function from it as: $\Psi=a_0(\tilde{\Psi})=r(\cos\theta)\left(e^{-\dfrac{r}{2a_0}}\right)\left(e^{-i\omega t}\right)$.

So that the invariant $ds^2$ turns out to be:

$$ds^2=a_0^{\,2}(\nabla\Psi)^2=\mathrm{Re}\left[\left(\dfrac{\partial\Psi}{\partial x_\mu}\right)\left(\dfrac{\partial\Psi}{\partial x_\nu}\right)\right]dx^\mu dx^\nu, \tag{34}$$

a multiple of square of the Bohr radii. Interestingly, the invariant $ds$ appearing as multiple of Bohr radii evoke a sense of aesthetics too, which one cannot but appreciate.

The metric coefficients corresponding to the quantum state $\Psi \equiv \Psi(r,\theta,t)$, are given by:

$$g_{rr} = \text{Re}\left[\left(\frac{\partial \Psi}{\partial r}\right)\left(\frac{\partial \Psi}{\partial r}\right)\right] = \left[\left(e^{-\frac{r}{a_0}}\right)\left(1 - \frac{r}{2a_0}\right)^2 \cos^2\theta \cos 2\omega t\right], \quad (35)$$

$$g_{\theta\theta} = \text{Re}\left[\left(\frac{\partial \Psi}{\partial \theta}\right)\left(\frac{\partial \Psi}{\partial \theta}\right)\right] = \left[\left(e^{-\frac{r}{a_0}}\right) r^2 \sin^2\theta \cos 2\omega t\right], \quad (36)$$

$$g_{tt} = \text{Re}\left[\left(\frac{\partial \Psi}{\partial t}\right)\left(\frac{\partial \Psi}{\partial t}\right)\right] = -\left[r^2 \omega^2 \left(e^{-\frac{r}{a_0}}\right) \cos^2\theta \cos 2\omega t\right]. \quad (37)$$

Similarly, we can also describe metric for the wave function $\Psi_{211}$ of the Hydrogen atom:

$$\Psi_{211} = \left(\frac{1}{4\sqrt{2\pi}}\right)\left(\frac{1}{a_0}\right)^{3/2}\left(\frac{r}{a_0}\right)\frac{e^{i\varphi}}{\sqrt{2}}(\sin\theta)\left(e^{-\frac{r}{2a_0}}\right)\left(e^{-i\frac{W_2}{\hbar}t}\right).$$

We follow the preceding example and consider the un-normalized dimensionless wave function corresponding to $\Psi_{211}$, and construct a wave function $\Psi$, from $\Psi_{211}$ as: $\Psi = \frac{\Psi_{211} + \Psi_{211}^*}{2}$, which is still a wave function of the Hydrogen atom. However, we choose to write the wave function $\Psi$ as $\Psi = \frac{\Psi_{211} + \Psi_{211}^*}{2}e^{-i\omega t}$, such that function $\Psi$ remains complex in nature. The metric coefficients corresponding to this quantum state $\Psi = r\sin\theta\left(e^{-\frac{r}{2a_0}}\right)\cos\varphi\left(e^{-i\omega t}\right)$, are given by:

$$g_{rr} = \text{Re}\left[\left(\frac{\partial \Psi}{\partial r}\right)\left(\frac{\partial \Psi}{\partial r}\right)\right] = \left[\sin^2\theta\cos^2\varphi\left(e^{-\frac{r}{a_0}}\right)\left(1 - \frac{r}{2a_0}\right)^2 \cos 2\omega t\right], \quad (38)$$

$$g_{\theta\theta} = \text{Re}\left[\left(\frac{\partial \Psi}{\partial \theta}\right)\left(\frac{\partial \Psi}{\partial \theta}\right)\right] = \left[r^2\cos^2\theta\cos^2\varphi\left(e^{-\frac{r}{a_0}}\right)\cos 2\omega t\right], \quad (39)$$

$$g_{\varphi\varphi} = \text{Re}\left[\left(\frac{\partial \Psi}{\partial \varphi}\right)\left(\frac{\partial \Psi}{\partial \varphi}\right)\right] = \left[r^2 \sin^2\theta \sin^2\varphi \left(e^{-\frac{r}{a_0}}\right) \cos 2\omega t\right], \tag{40}$$

$$g_{tt} = \text{Re}\left[\left(\frac{\partial \Psi}{\partial t}\right)\left(\frac{\partial \Psi}{\partial t}\right)\right] = -\left[\omega^2 r^2 \sin^2\theta \cos^2\varphi \left(e^{-\frac{r}{a_0}}\right) \cos 2\omega t\right]. \tag{41}$$

This is metric with the signature (+, +, +, -) and the corresponding line element appears as:

$$ds^2 = a_0^2 \, \text{Re}(\nabla\Psi)^2 = \left[\sin^2\theta \cos^2\varphi \left(e^{-\frac{r}{a_0}}\right)\left(1 - \frac{r}{2a_0}\right)^2 \cos 2\omega t\right] dr^2 + \left[r^2 \cos^2\theta \cos^2\varphi \left(e^{-\frac{r}{a_0}}\right) \cos 2\omega t\right] d\theta^2$$

$$+ \left[r^2 \sin^2\theta \sin^2\varphi \left(e^{-\frac{r}{a_0}}\right) \cos 2\omega t\right] d\varphi^2 - \left[\omega^2 r^2 \sin^2\theta \cos^2\varphi \left(e^{-\frac{r}{a_0}}\right) \cos 2\omega t\right] dt^2$$

$$\tag{42}$$

We notice that the wave function $\Psi$ of the Hydrogen atom defined here, admits metric in the four space with co-ordinates $(r, \theta, \varphi, t)$, whereas $\Psi_{100}, \Psi_{200}$ and $\Psi_{210}$ fail to do so.

*Invariance under the Lorentz' (relativistic) transformations*

As discussed, the metric structure in the configuration space is invariant under the coordinate transformations only. But, if the wave function under consideration is relativistic, the invariance of the metric under the Lorentz' transformations is also ensured‡. However, the term 'Lorentz' invariance is a misnomer in this context. As the relativistic wave function of the Hydrogen atom, was given by Dirac. Thus we simply mean that a relativistic formulation turns non-relativistic in the given limits. In such a case the metric in the configuration space is invariant under the coordinate transformation as well as relativistic transformation of the wave function. We now illustrate an example of the metric corresponding to relativistic wave function of Hydrogen like atom.

The relativistic wave function of Hydrogen like atom as proposed by Dirac [14] (Bjorken *et al*. 1965) can be given as:

$$\tilde{\Psi}_{n=1, j=1/2, \uparrow}(r,\theta,\varphi) = \frac{(2mZ\alpha)^{3/2}}{\sqrt{4\pi}} \sqrt{\frac{1+\gamma}{2\Gamma(1+2\gamma)}} (2mZ\alpha r)^{\gamma-1} e^{-mZ\alpha r} \left(\frac{i(1-\gamma)}{Z\alpha}\right) \sin\theta e^{-i\varphi}; \quad (43)$$

where $\gamma \cong \sqrt{1 - \frac{(Z\alpha)^2}{n}}$, $\alpha = \frac{e^2}{\hbar c}$ is fine structure constant, and $\frac{1}{2mZ\alpha} \sim a_0$ is the Bohr radii. In the non-relativistic limits $\gamma \to 1$ and $\frac{(1-\gamma)}{Z\alpha} \to 0$, and the wave function reduces to Schrödinger wave function. By taking the un-normalized wave function and transferring the other constants to left hand side, the above wave function reduces to a simpler form: $\tilde{\Psi} = i\left(\frac{r}{a_0}\right)^{\gamma-1} e^{-\frac{r}{a_0}} \sin\theta e^{-i\varphi}$. We now construct a wave function $\Psi$, from $\tilde{\Psi}$ as: $\Psi = \frac{\tilde{\Psi} + \tilde{\Psi}^*}{2}$, and choose to write the wave function $\Psi$ as $\Psi = \frac{\tilde{\Psi} + \tilde{\Psi}^*}{2} e^{-i\omega t}$, such that function $\Psi$ remains complex in nature. Now, with the help of the wave function

$\Psi = (r)^{\gamma-1} e^{-\frac{r}{2a_0}} \sin\theta \sin\varphi e^{-i\omega t}$; we formulate our metric as:

$ds^2 = a_0^{2(\gamma-1)} \{\text{Re}(\nabla\Psi)^2\} = \text{Real Part}\left[(\nabla_\mu \Psi)(\nabla_\nu \Psi)\right] dx^\mu dx^\nu$, or

$$ds^2 = \left[r^{2(\gamma-2)}\left(e^{-\frac{r}{a_0}}\right)\sin^2\theta\sin^2\varphi\left(\gamma-1-\frac{r}{2a_0}\right)^2 \cos 2\omega t\right] dr^2 + \left[r^{2(\gamma-1)}\left(e^{-\frac{r}{a_0}}\right)\cos^2\theta\sin^2\varphi\cos 2\omega t\right] d\theta^2$$

$$+ \left[r^{2(\gamma-1)}\left(e^{-\frac{r}{a_0}}\right)\sin^2\theta\cos^2\varphi\cos 2\omega t\right] d\varphi^2 - \left[r^{2(\gamma-1)}\left(e^{-\frac{r}{a_0}}\right)\sin^2\theta\sin^2\varphi\cos 2\omega t\right] dt^2.$$

(44)

Thus it is quite evident that contrary to the common perception that a metric for quantum state can yield a natural metric in the configuration space only when the limit $\hbar \to 0$, we find the metric of quantum states in the configuration space without imposing the limiting condition $\hbar \to 0$. The Planck's constant $\hbar$ is absorbed in the quantity like Bohr radii $\frac{1}{2mZ\alpha} \sim a_0$. Also, we find that the metric in the configuration space could turn out to be a metric of space-time, wherever configuration space coincides with space-time (see ref: 2-3). This is with assumption that wherever the configuration space coincides with *space-time*, the natural metric on $CP(N)$ in the $\hbar \to 0$ limit gives a spatial metric [2-5].

Hydrogen atom represents the matter in its simplest form. Therefore, the investigation of the geometric features associated with Hydrogen atom has a rationale behind it.

## 4. SUMMARY AND DISCUSSION

This paper aims to present a discussion on the metric of quantum states in a comprehensive perspective. Interestingly, the metric of quantum state space explored in the geometric quantum mechanics, has gained renewed interest of scientific community as formalism pertaining to background independent quantum mechanics (BIQM). We strongly push our demand that the configuration space metric can be the actual physical *spatial* metric in special cases. The suitable quantum system can then have a very special configuration space and should describe gravity in its premise.

We in this paper have further explored the reasons of invariance of the geometric structure like metric in the ray space. Also, it is interesting to see that the mechanism causing invariance under the local gauge transformations plays important role in the construction of 'quantum information theory' [15, 16].

This discussion summarizes here the metric structures so far explored in the geometric quantum mechanics. We have encountered metric structures on three different manifolds: Kähler manifold or $CP(N)$, Riemannian manifold, and space-time (pseudo-Riemannian) manifold.

If the metric of quantum states is defined with the complex coordinates in the quantum state space, known as Fubini-Study metric, it lies on the Kähler manifold or $CP(N)$, which is identified with the quotient set $\frac{U(N+1)}{U(N) \times U(1)}$. By insisting on the diffeomorphism invariance in the state space and on preserving the desirable complex projective properties of Cartan's rank 1 symmetric spaces such as $CP(N)$, an extended framework for such a representation has been suggested as the Grassmannian: $Gr(C^{n+1}) = Diff(C^{n+1})/Diff(C^{n+1}, C^n \times \{0\})$.

Apart from the fundamental difference that, the metric of quantum state space is metric in the ray space and the metric otherwise stated is in the configuration space, there are many other differences, including the underlying difference in the signature of the metric structures. The signature of the metric of quantum state space is always positive definite. Where as, the metric in the configuration space need not be positive definite, as it is clear from the examples cited in this discussion.

And if the metric of quantum states is defined with local co-ordinates that are not complex, it lies on the base manifold with Riemannian character, and the local gauge group $GL(n, R)$ is also admissible.

Whereas, if the metric of quantum states is defined in the configuration space with the space-time co-ordinates, the base manifold **M** on which it lies, carries a (pseudo)

Riemannian metric as well, and the tetrad can naturally be chosen to bring the metric $g_{\mu\nu}$ to a diagonal Minkowski form. And then the Lorentz group $SO(3,1)$ could also appear as a local gauge group.

We must notice that the group symmetry observed in the quotient set $\frac{U(N+1)}{U(N)\times U(1)}$ in case of Fubini-Study metric is the symmetry over the transformations of the wave functions. Whereas, the group symmetry mentioned in the later cases as $GL(n,R)$ and $SO(3,1)$, if observed, could be due to the transformations of co-ordinates.

On a Kähler manifold in the quantum state space, invariance under the local gauge transformations is same as invariance under the coordinate transformations. This is due to the fact that in the quantum state space, quantum states themselves could play the role of coordinates. On the other hand, a metric with classical nature does not admit invariance under the local gauge transformations. And for which, the invariance under the coordinate transformations is enough.

Thus, we find that the metric in the configuration space has lost invariance under local gauge transformations, but it is still invariant at least under the coordinate transformations. Also, if the wave function subject to condition is relativistic, it is invariant under the Lorentz' transformation as well.

Among other distinctions, we find that the metric coefficients $g_{\mu\nu} = [\langle \partial_\mu \Psi | \partial_\nu \Psi \rangle - \langle \partial_\mu \Psi | \Psi \rangle \langle \Psi | \partial_\nu \Psi \rangle]$, defined in the metric of quantum state space, are under the integrals and therefore constant. Where as, the metric coefficients in the case of metric in the configuration space are not constant.

Since, the metric coefficients in the metric of quantum state space are constant, all their derivatives readily vanish. Consequently, one cannot calculate Christoffel symbols, Ricci tensor, and Einstein tensor. Where as, for the metric of quantum states in the configuration space, there is possibility that one can explore the other geometric features associated with the metric of quantum states.

If we insist on the desired relation between the quantum state space metric and an arbitrary metric on the classical configuration space, then the kinematics of QM has to be altered [2-4]. Moreover, if the induced classical configuration space is to be actual space of space-time, only a special quantum system will do. We are thus induced to explore an appropriate metric arising due to quantum states and living on the space-time manifold, which in turn may enable us to do general relativity (GR) on it.

## Acknowledgement

The author wishes to express his gratefulness to Prof. A. Ashtekar and late J. Anandan for explaining the niceties of the geometric quantum mechanics.

‡We ought to call the invariance resulting from the use of relativistic wave function as Dirac's invariance, instead of Lorentz' invariance, as the relativistic wave function of the Hydrogen atom was given by Dirac.